\documentclass{ws-procs961x669}            

\usepackage{mathrsfs}
\usepackage{amsfonts}
\usepackage{amssymb}
\usepackage{bbm}

\DeclareMathOperator{\supp}{supp}
\def\fmax{\phi_{\text{max}}}
\newcommand{\nord}[1]{{:}#1{:}}
\newcommand{\coin}[1]{\left[\!\!\left[#1\right]\!\!\right]}

\newcommand{\be}{\begin{equation}}
	\newcommand{\ee}{\end{equation}}
\newcommand{\bea}{\begin{eqnarray}}
	\newcommand{\eea}{\end{eqnarray}}
\def\bml{\begin{subequations}}
	\def\blea{\bml\begin{eqnarray}}
		\def\eml{\end{subequations}}
	\def\elea{\end{eqnarray}\eml}

\begin{document}
\title{Classical and quantum strong energy inequalities and the Hawking singularity theorem\footnote{Contributed talk by E.-A.K. in the AT3 session on Wormholes, Energy Conditions and Time Machines at the Fifteenth Marcel Grossmann Meeting.}}

\author{P. J. Brown, C. J. Fewster and E.-A. Kontou$^*$}

\address{Department of Mathematics, University of York,\\
Heslington, York, YO10 5DD, UK\\
$^*$E-mail: eleni.kontou@york.ac.uk}

\begin{abstract}
Hawking's singularity theorem concerns matter obeying the strong energy condition (SEC), which means that all observers experience a non-negative effective energy density (EED). The SEC ensures the timelike convergence property. However, for both classical and quantum fields, violations of the SEC can be observed even in the simplest of cases, like the Klein-Gordon field. Therefore there is a need to develop theorems with weaker restrictions, namely energy conditions averaged over an entire geodesic and weighted local averages of energy densities such as quantum energy inequalities (QEIs). We present lower bounds of the EED for both classical and quantum scalar fields allowing nonzero mass and nonminimal coupling to the scalar curvature. In the quantum case these bounds take the form of a set of state-dependent QEIs valid for the class of Hadamard states. We also discuss how these lower bounds are applied to prove Hawking-type singularity theorems asserting that, along with sufficient initial contraction, the spacetime is future timelike geodesically incomplete.
\end{abstract}


\bodymatter

\section{Introduction}\label{sec:intro}

A spacetime is defined to be singular if it possesses at least one incomplete geodesic. The question of whether or not cosmological models either originate or terminate in singularities has been an active subject of research since the formulation of the general theory of relativity. Initial efforts focused on models with high levels of symmetry, until Raychaudhuri's 1955 paper \cite{Raychaudhuri:1953yv} paved the way to more general results. The Raychaudhuri equations in their modern form~\cite{Ehlers:1993} present the evolution of geodesic congruences, and are the heart of most singularity theorems. For a timelike irrotational congruence with velocity field $U^\mu$, the expansion $\theta=\nabla \cdot U$ satisfies
\begin{equation}
\label{eqn:ray0}
\nabla_U\theta =R_{\mu \nu} U^\mu U^\nu-2\sigma^2 -\theta^2/(n-1)  \,,
\end{equation}
where $n$ is the spacetime dimension, $\sigma$ is the shear scalar and $R_{\mu\nu}$ is the Ricci tensor. Senovilla~\cite{Senovilla:2014} has described the singularity theorems in terms of a `pattern theorem' with three ingredients: an energy condition establishes a focussing effect for geodesics, a causality condition removes the possibility of closed timelike curves, and a boundary or initial condition establishes the existence of some trapped region of spacetime. We divide singularity theorems into `Hawking-type' after Hawking's original theorem~\cite{Hawking:1966sx} and `Penrose-type' after Penrose's~\cite{Penrose_prl:1965}, depending on whether they demonstrate timelike or null geodesic incompleteness respectively. Hawking-type results, which concern us here, are based on the strong energy condition (SEC), which requires that the effective energy density (EED) is non-negative. The EED is defined as $\rho_U := T_{\mu \nu} U^\mu U^\nu-T/(n-2)$ and is easily made negative at individual points even for the classical minimally coupled scalar field. This situation is exacerbated in quantum field theory (QFT), in which none of the pointwise energy conditions can hold~\cite{Epstein:1965zza}. 

For these reasons there has long been interest in establishing singularity theorems under weakened energy assumptions, such as requiring the averaged energy along the geodesic to be non-negative. Examples of singularity theorems with such conditions include Refs.~\citenum{Tipler:1978zz,ChiconeEhrlich:1980,Borde:1987b} but none of them address the case of a condition obeyed by quantum fields. Within a QFT the weakened energy conditions take the form of quantum energy inequalities (QEIs). These were introduced by Ford~\cite{Ford:1978qya} and express a restriction on the possible magnitude and duration of any negative energy densities or fluxes. Our approach follows Ref.~\citenum{Fewster:2010gm} which proved singularity theorems with energy conditions inspired by QEIs. For the Hawking-type theorem, it is assumed that
\begin{equation}
	\label{eqn:energyfg}
\int R_{\mu \nu} \dot{\gamma}^\mu \dot{\gamma}^\nu f^2(\tau) d\tau \geq -|||f|||^2 \,,
\end{equation}
on any timelike geodesic. Here $\dot{\gamma}^\mu$ is the tangent to the timelike geodesic $\gamma$, $||| \cdot |||^2$ is a Sobolev norm of the form $|||f|||^2=\sum_{\ell=0}^L Q_\ell ||f^{(\ell)}||^2$ and $Q_\ell$ are non-negative constants. It is also assumed that there is a compact Cauchy hypersurface intersected by $\gamma$ at $\tau=0$. Then the spacetime is future timelike geodesic incomplete provided that the required initial contraction obeys
\begin{equation}
\theta(0) \leq -\frac{c}{2}-\int_{-\tau_0}^\infty R_{\mu \nu} \dot{\gamma}^\mu \dot{\gamma}^\nu d\tau -|||f|||^2 \,,
\end{equation}
for some $c>0$ and $f\in C^\infty (\mathbb{R})$ with $\supp{f} \subset [-\tau_0,\infty)$, $f(\tau)=e^{-c\tau/(n-1)}$ on $[0,\infty)$. It might seem strange at first that large positive values of $R_{\mu \nu} \dot{\gamma}^\mu \dot{\gamma}^\nu$ require larger initial contraction. The reason is that the average energy condition of Eq.~\eqref{eqn:energyfg} means that large positive energies in the past (before the $\theta(0)$ is measured) allow large negative energies in the future that can lead to a divergence of the congruence. 

The purpose of this contribution is to describe bounds on the EED for classical and quantum fields, show that they are of the form of Eq.~\eqref{eqn:energyfg} and estimate the contraction needed to prove singularity theorems. In Sec.~\ref{sec:class} we treat the classical Einstein-Klein-Gordon field while in Sec.~\ref{sec:QSEI} we present a bound on the renormalised expectation value of the EED for the non-minimally coupled quantum scalar field. In Sec.~\ref{sec:conc} we conclude with a summary and discussion of future work. Our geometrical conventions are those of Ref.~\citenum{Fewster:2010gm} and we work on a globally hyperbolic manifold $M$.

\section{A singularity theorem for the Einstein-Klein-Gordon field}
\label{sec:class}

The non-minimally coupled scalar field obeys the field equation $P_\xi\phi = 0$ with  $P_\xi:=\Box_g+m^2+\xi R$, where $\xi$ is the coupling constant and 
\begin{equation}
\label{eqn:tmunu}
T_{\mu \nu}=(\nabla_\mu \phi)(\nabla_\nu \phi)+\frac{1}{2} g_{\mu \nu} (m^2 \phi^2-(\nabla \phi)^2)+\xi(g_{\mu \nu} \Box_g-\nabla_\mu \nabla_\nu-G_{\mu \nu}) \phi^2 \,.
\end{equation}
For this stress-energy tensor, after some manipulation, the averaged EED on a timelike geodesic $\gamma$ is
\bea
&&\int_\gamma d\tau \,\rho_U \, f^2(\tau)=\int_\gamma d\tau \bigg\{-\frac{1-2\xi}{n-2} m^2 f^2(\tau)+\left(1-2\xi\frac{n-1}{n-2}\right) (\nabla_{\dot{\gamma}} \phi)^2f^2(\tau) \nonumber \\
&&+\frac{2\xi}{n-2} h^{\mu \nu} (\nabla_\mu \phi) (\nabla_\nu \phi)f^2(\tau) +2\xi  [\nabla_{\dot{\gamma}} (f(\tau))\phi]^2 -2\xi \phi^2 (f'(\tau))^2-\xi \mathcal{R}_\xi \phi^2 \bigg\} \,,
\eea
where $\mathcal{R}_\xi=R_{\mu \nu} \dot{\gamma}^\mu \dot{\gamma}^\nu-2\xi/(n-2) R$, and
$h^{\mu \nu}=\dot{\gamma}^\mu \dot{\gamma}^\nu-g^{\mu \nu}$ is a positive definite metric. For $\xi \in [0,\xi_c]$, where $\xi_c$ is the conformal coupling constant, all the curvature independent terms have a definite sign, so
\bea
\label{eqn:cline}
\int_\gamma d\tau \, \rho_U \, f^2(\tau)&\geq& - \int_\gamma d\tau \bigg\{ \frac{1-2\xi}{n-2} m^2 f^2(\tau)+\xi \bigg(2( f'(\tau))^2+\mathcal{R}_\xi f^2(\tau) \bigg) \bigg\} \phi^2 \,.
\eea
For a field obeying the Einstein's equation, we have that $8\pi \rho_U=R_{\mu \nu} \dot{\gamma}^\mu \dot{\gamma}^\nu$ and $(n/2-1)R=8\pi T$. Moving the curvature terms to the left hand side of Eq.~\eqref{eqn:cline}
\be
\label{eqn:boundcl}
\int R_{\mu \nu} \dot{\gamma}^\mu \dot{\gamma}^\nu f(\tau)^2\,d\tau \geq -Q(\|f'\|^2+ \tilde{Q}^2\|f\|^2),
\ee
with $Q$ and $\tilde{Q}$ positive constants that depend on $\fmax \geq |\phi| $, $m$, $\xi$ and $n$. We can show that if
\bea
	&&\theta(0) <-\tilde{Q}\sqrt{Q(n-1)+\frac{Q^2}{2}}-\frac{1}{2}Q K \coth{(K\tau_0)}\,, \nonumber\\
	 && \qquad \qquad \text{ with } K^2 \geq \tilde{Q}^2+Q^{-1} R_{\mu \nu} \dot{\gamma}^\mu \dot{\gamma}^\nu \text{ on } (-\tau_0,0] \,,
\eea
on a compact Cauchy surface, the spacetime is future timelike geodesic incomplete.

Finally, we want to estimate the necessary contraction for realistic scalar fields, using a hybrid model: a quantised scalar field in Minkowski spacetime of dimension $4$, in a thermal state of temperature $T<T_m$, $T_m=mc^2/k$ with the maximum field value set by the expectation value of the Wick square at that temperature, $\fmax^2\sim\langle {:}\phi^2{:}\rangle_T $. As an example we consider the Higgs field with $m=125\textrm{GeV}/c^2$, $\theta(0) \sim 10^{-14}\textrm{s}^{-1}$ for temperature up to $T=10^{13}\textrm{K}$. Similar results are obtained for other examples. We conclude that when the field mass is taken equal to that of an elementary particle, and provided that the temperature stays below early universe levels we need very little initial contraction for geodesic incompleteness \footnote{A detailed proof of the theorem and estimation of the initial contraction is given in Ref.~\citenum{Brown:2018hym}.}.

\section{Strong quantum energy inequality}
\label{sec:QSEI}

To quantise the EED we will follow the algebraic approach. We only consider quasifree, Hadamard states $\omega$, for which the two-point function $W(x,y) = \left<\Phi(x)\Phi(y)\right>_{\omega}$ has a prescribed singularity structure so that the difference between two states is smooth. 

We follow the methods of Hollands and Wald~\cite{Hollands:2001nf,Hollands:2004yh} to quantize the stress-energy tensor in a systematic and locally covariant way, and then define the EED as a quantum field by $\rho_U(f)=T_{\mu\nu}(U^\mu U^\nu -g^{\mu\nu}/(n-2)f)$. We are interested in expectation values of the quantized EED in state $\omega$, normal ordered relative to a reference Hadamard state $\omega_0$, $\langle {:}\rho_U{:} (f)\rangle_{\omega} = \langle  \rho_U (f)\rangle_{\omega} - \langle  \rho_U (f)\rangle_{\omega_0}$.

We now turn to the derivation of a quantum strong energy inequality (QSEI), a bound on the renormalised expectation value of EED averaged along a timelike geodesic $\gamma$. Choose any smooth $n$-bein $e_a$ ($a=0,\ldots,n-1)$ on a tubular neighbourhood of $\gamma$, so that $U^\mu=e_0^\mu$ is everywhere timelike and agrees with $\dot{\gamma}^\mu$ on $\gamma$. The expectation values of the EED in Hadamard state $\omega$ can be written in terms of the coincidence limits, $\coin{\cdot}$ acting on $\nord{W}=W-W_0$. Then we have 
\be
\langle \nord{\rho_{U}}\rangle_\omega =\coin{ \hat{\rho}_1 \nord{W} } +\coin{ \hat{\rho}_2 \nord{W}}  + \left(\xi\mathcal{R}_\xi-\frac{1-2\xi}{n-2} m^2 \right) \coin{\nord{W}} \text{ with}
\ee
\blea
\hat{\rho}_1 &=&\left( 1-2\xi \frac{n-1}{n-2} \right) (\nabla_{U} \otimes \nabla_{U} )+\frac{2\xi}{n-2} \sum_{a=1}^{n-1} (\nabla_{e_a} \otimes \nabla_{e_a})  \,, \\
\hat{\rho}_2 &=& -2\xi (\mathbbm{1} \otimes_{\mathfrak{s}} U^\mu U^\nu \nabla_\mu \nabla_\nu ) \,,
\elea
where $\otimes_{\mathfrak{s}}$ is the symmetrised tensor product $P \otimes_{\mathfrak{s}} P'=[ (P \otimes P')+(P' \otimes P)]/2$. The contribution of the terms deriving from $\hat{\rho}_1$ to the averaged EED can be bounded from below, uniformly in $\omega$, using the methods of Ref.~\citenum{Fewster:1999gj}. By contrast, the mass term is negative definite for $\xi<1/2$, while the geometric term $\mathcal{R}_\xi$ has no definite sign in general. This leaves $\hat{\rho}_2$, the contribution of which can be manipulated to a more convenient form. With these considerations we can prove 
\be \label{eqn:qboundline}
\langle \nord{\rho_U} \circ \gamma \rangle_\omega (f^2) \geq - \left( \mathfrak{Q}_1[f]  +\langle \nord{\Phi^2} \circ \gamma \rangle_\omega (\mathfrak{Q}_2[f]+\mathfrak{Q}_3[f])\right)  \text{ where}
\ee
\blea
\mathfrak{Q}_1[f]&=&\int_0^\infty \frac{d\alpha}{\pi} \left( \phi^*(\hat{\rho}_1 \, W_0)(\bar{f_\alpha},f_\alpha)+2\xi \alpha^2  \phi^* W_0  (\bar{f_\alpha},f_\alpha) \right) \,, \\
\mathfrak{Q}_2[f]&=&\frac{1-2\xi}{n-2}m^2 f^2(\tau) +2\xi  ( f'(\tau))^2 \text{ and } \mathfrak{Q}_3 [f]=\xi  \mathcal{R}_\xi(\gamma(\tau)) f(\tau)^2\,.
\elea
Here $\phi^*$ is the distributional pull-back on the geodesic. An important feature of the QEI~\eqref{eqn:qboundline}, is that the lower bound depends on the state of interest $\omega$. We should note however, that the only nontrivial quantum field appearing in the bound is the Wick square $\nord{\Phi^2}$ which enables us to show that the QEI derived is nontrivial (see Ref.~\citenum{Fewster:2018pey} for more details). 

The establishment of a QSEI is a first step towards a Hawking-type singularity theorem result employing QEI hypotheses. Here we outline a method of obtaining such a result, part of an ongoing work to appear elsewhere. The singularity theorems require a geometric assumption, a condition on the curvature rather than the stress-energy tensor. In the case of classical fields we can use the Einstein equation while when we are treating quantum fields on a classical curved background we can instead use the semiclassical Einstein equation (SEE), $\langle T_{\mu \nu}  \rangle_\omega =8\pi G_{\mu \nu}$. In order to use the QSEI and the SEE for general curved spacetimes the EED needs to be renormalised by subtracting the Hadamard parametrix instead of a reference state. To overcome this problem we note there is evidence (see e.g. Ref.~\citenum{Kontou:2014tha}) that in situations where the curvature is bounded we can find a uniform length $\tau_0$ which is small compared to local curvature. Then the Hadamard parametrix on $\tau_0$ approximates that of flat spacetime. Following, we can use Eq.~\eqref{eqn:qboundline} for $f \in C_0^\infty$ supported only to intervals with lengths at most $2\tau_0$. Choosing a state $\omega$ and a metric $g_{\mu \nu}$ that satisfy the EED and the stress energy tensor renormalised with respect to the Minkowski vacuum,
\be
\label{eqn:rmunueach}
\int d\tau f^2(\tau) R_{\mu \nu} U^\mu U^\nu \geq -8\pi \bigg[ \int_0^\infty \frac{d\alpha}{\pi} \phi^*((\nabla_{U} \otimes \nabla_{U}) \, W_0)(\bar{f_\alpha},f_\alpha) +\frac{\mu^2 \fmax^2||f||^2}{n-2} \bigg] \,,
\ee
where we also set $\xi=0$, $\mu$ is the mass of the field and we restrict to a class of Hadamard states for which $|( \colon \Phi^2 \colon \gamma)_\omega| \leq \fmax^2$. 

To discuss averages over long timescales we will use a partition of unity. We define bump functions each supported on $(-\tau_0,\tau_0)$ and thus obtain a sum of integrals, each of which can be bounded by Eq.~\eqref{eqn:rmunueach}. Then for even numbers of spacetime dimensions ($m=n/2$), we can show that for $f$ supported on $(-\infty, \infty)$ we have
\be
\int_{-\infty}^\infty d\tau f^2(\tau) R_{\mu \nu} U^\mu U^\nu \geq -Q_m(||f^{(m)}||^2+\tilde{Q}_m^2||f||^2)  \,,
\ee
where the $Q_m$ and $\tilde{Q}_m$ depend on $\fmax$, $\mu$, and the maximum values of the chosen bump function and its derivatives. This is an inequality of the form of Eq.~\eqref{eqn:energyfg} so it can be used as a condition to a Hawking-type singularity theorem.

\section{Conclusions}
\label{sec:conc}

In this work we presented the derivation of lower bounds on EED for the classical and quantum non-minimally coupled scalar field. In the classical case we presented the proof of a Hawking-type singularity theorem for the Einstein-Klein-Gordon field and estimated the required initial contraction to have future timelike geodesic incompleteness. In the quantum case, using the SEE we showed that we can obtain a condition on the curvature to be used as a hypothesis for a singularity theorem. 

Apart from estimating the required initial contraction for the quantum case, the obvious extension of this work is the derivation of an absolute (Hadamard renormalised) QSEI for spacetimes with curvature. Another direction is the derivation of a Penrose-type singularity theorem with a condition obeyed by quantum fields. Finally we should note that similar proofs with weakened energy conditions seem to be possible for other classical relativity theorems. 

\section*{Acknowledgements}

This work is part of a project that has received funding from the European Union's Horizon 2020 research and innovation programme under the Marie Sk\l odowska-Curie grant agreement No. 744037 ``QuEST''.

\bibliographystyle{ws-procs961x669}
\bibliography{paper}

\end{document}